\documentclass[10pt]{article}
\usepackage{amssymb,verbatim,epsfig}
\usepackage{amsmath,setspace}
\usepackage[usenames,dvipsnames]{color}
\usepackage{url}
\usepackage{amssymb, amsmath, verbatim, epsfig}
\usepackage{setspace}
\usepackage{lineno}
\usepackage{hyperref}

\newcommand{\fig}[1]{{\bf Figure \ref{#1}}}

\usepackage{natbib}
\setcitestyle{numbers} 
\usepackage{caption}
\usepackage{enumitem}
\usepackage{sidecap}
\title{\bf 
\vspace{-1in}
Social Insects and Beyond: The Physics of Soft, Dense Invertebrate Aggregations
\author{Olga Shishkov$^{1}$, Orit Peleg$^{1,2,3*}$ \\	
\text{\small{Biofrontiers Institute$^1$, and Department of Computer Science$^2$}} \\
 \text{\small{University of Colorado Boulder, Boulder, CO 80309, USA}} \\
 \text{\small{Santa Fe Institute$^3$, Santa Fe, NM 87501, USA}} \\
 \text{\small{orit.peleg@colorado.edu$^*$}} \\
\date{}
}}
\begin{document}
\maketitle

\section*{Abstract}

Aggregation is a common behavior by which groups of organisms arrange into cohesive groups. Whether suspended in the air (like honey bee clusters), built on the ground (such as army ant bridges), or immersed in water (such as sludge worm blobs), these collectives serve a multitude of biological functions, from protection against predation to the ability to maintain a relatively desirable local environment despite a variable ambient environment. In this review, we survey dense aggregations of a variety of insects, other arthropods, and worms from a soft matter standpoint. An aggregation can be orders of magnitude larger than its individual organisms, consisting of tens to hundreds of thousands of individuals, and yet functions as a coherent entity. Understanding how aggregating organisms coordinate with one another to form a superorganism requires an interdisciplinary approach. We discuss how the physics of the aggregation can yield additional insights to those gained from ecological and physiological considerations, given that the aggregating individuals exchange information, energy, and matter continually with the environment and one another. While the connection between animal aggregations and the physics of non-living materials has been proposed since the early 1900s, the recent advent of physics of behavior studies provides new insights into social interactions governed by physical principles. Current efforts focus on eusocial insects; however, we show that these may just be the tip of an iceberg of superorganisms that take advantage of physical interactions and simple behavioral rules to adapt to changing environments. By bringing attention to a wide range of invertebrate aggregations, we wish to inspire a new generation of scientists to explore collective dynamics and bring a deeper understanding of the physics of dense living aggregations.

\section{Introduction}

Collective phenomena exist on a multitude of scales of biology, from cells that assemble into organs \cite{aman2010cell, trepat2009physical} to insects that assemble into bivouacs \cite{kronauer2020army,vernerey2019biological} to humans who assemble into dense concert audiences \cite{silverberg2013collective}. These aggregations perform a variety of functions to benefit their constituent organisms, including maintaining a desirable internal environment despite variable ambient conditions, enhancing locomotion, and avoiding predation \cite{parrish1999complexity, moussaid2009collective, camazine2001self,sumpter2006principles}. The oldest documented animal aggregation, a chain of arthropods, appeared five hundred million years ago during the Cambrian period \cite{hou2008collective, blazejowski2016ancient,vannier2019collective}. As numerous vertebrate and invertebrate animals evolved over the millennia, aggregations with a variety of shapes, sizes, and behaviors developed \cite{chandra2021colony}. Understanding how these aggregations coalesce and behave is an active field of study. 

As it would take much more than a single review to discuss all of the collectives of arthropods, worms, fish, birds, and mammals, we limit our focus to ``dense'' aggregations of invertebrates - where the bodies of the individuals make physical contact with one another. In these aggregations, the members can be tightly packed (e.g., western honey bees forming leg--leg bonds in a swarm \cite{peleg2018collective}) or just barely touching (e.g., whirligig beetles gathering on the water surface \cite{voise2011capillary}). In contrast, individuals in sparse aggregations do not physically touch each other and are separated by air or water, such as midges in swarms and birds in  flocks \cite{van2020environmental,gray1991fast}. 

\begin{SCfigure}
		\includegraphics[width = 4.5in,keepaspectratio=true]
		{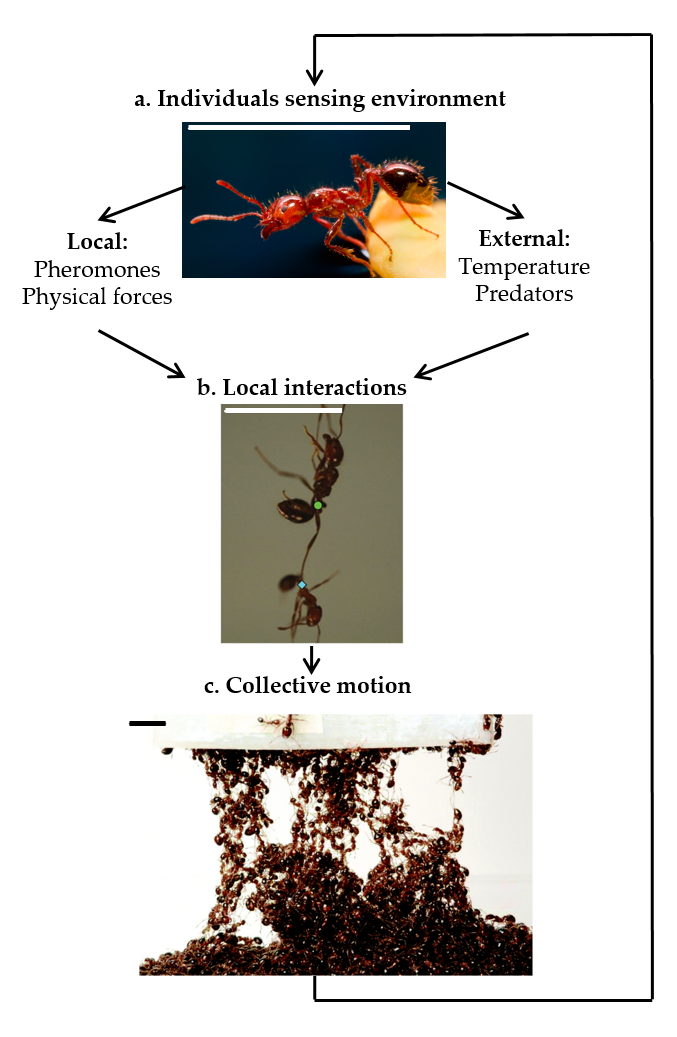}
	\caption{Flowchart indicating coupling between the individual organisms, the superorganism, and environment, using fire ants (\textit{Solenopsis invicta}) as a paradigm. Individuals (a) sense the local and external environment. They react to it by, for example, pulling on one another through the connection in their legs (b). This results in a collective response in which fire ants resist the aggregation being pulled apart by forming chains of ants pulling downwards (c). The individuals in the aggregation continue sensing mechanical forces and reacting to them until the aggregation breaks or reaches mechanical stability \cite{foster2014fire,phonekeo2016ant}. Photograph in (a) by Tim Nowack and David L. Hu. All scale bars are 0.5 cm long.}
	\label{flowchart}
\end{SCfigure}

Invertebrate aggregations span a wide range of length scales and consist of dozens to hundreds of thousands of individuals. Hence, an individual inside an aggregation may exchange information with its local environment and neighboring individuals, but it cannot directly interact with individuals far from it in the aggregation. The local response of individuals may propagate within the aggregation, activating dynamical processes (\ref{g:dynamical}) inside the group and leading to the collective motion of the entire aggregation. Consequently, this emergent global response leads to changes in the information perceived locally by individuals \cite{parrish1999complexity, moussaid2009collective}. This feedback loop is a central feature of dense aggregations, as illustrated in \fig{flowchart} using fire ants as a paradigm. 

An aggregation of densely packed invertebrates can have properties of both solid and liquid materials (\ref{g:soft}). The analogy between aggregations and viscoelastic (\ref{g:viscoelasticity}) materials is well demonstrated with ant aggregations: linked Argentine ants flow out of a faucet under the force of gravity  (\ref{g:viscosity}) \cite{bonabeau1998dripping}, and a ball of fire ants expands after compression like an elastic material (\ref{g:elasticity}) \cite{tennenbaum2016mechanics}. This connection between animal aggregations and non-living soft materials has been proposed throughout the 1900s. In 1931, W. C. Allee wrote about the many different types of animal aggregations and compared swarms of flying insects to a group of particles undergoing Brownian motion  (\ref{g:brownian})\cite{allee1931animal}. In 1978, Oster and Wilson hypothesized a connection between liquid flows and social insect behavior \cite{oster1978caste}. Since then, investigating the analogy between living aggregations and non-living materials using techniques from soft matter, fluid mechanics, and medical physics has increased in popularity for understanding how aggregations form and behave \cite{camazine2001self,sumpter2006principles}.

In the next sections, we describe the work that has been done on understanding invertebrates from a physics perspective. We describe the functions of aggregations in Section 2. We then categorize a variety of species that are known to aggregate by the physical properties of individuals in Section 3 and categorize the resulting aggregations by their material properties in Section 4. We then present the current literature on the analogies between aggregations as materials and techniques to study their motility in Section 5 and discuss potential areas for further research in Section 6. Throughout the text, the reader can refer to the glossary in Section \ref{Glossary} for definitions of physics terminology. 

\section{The functions and benefits of aggregations}
In this review, we describe a variety of candidate invertebrate aggregations for physics-based investigation. Out of the countless aggregations that can be found in entomological journal articles and books such as \cite{holldobler2009superorganism} and \cite{costa2006other}, we select sample aggregations that are representative of a variety of resulting physical structures. Unlike liquid crystals (\ref{g:LC}) or entangled polymers (\ref{g:entanglement}, \ref{g:polymer}), aggregations of invertebrates usually form to benefit the constituent individuals. Before delving into the physical properties of these aggregations, we must first understand the functions performed by the aggregations.

Some aggregations are long-term structures that house an entire invertebrate colony. Individuals may leave or join, yet the aggregation remains in place. For instance, giant honey bees build their permanent wax nests suspended outside and covered in layers of bees \cite{woyke2016shape, kastberger2008social}. Similarly, army ants form a bivouac by clinging together with their legs when they are not marching \cite{kronauer2020army, rettenmeyer2011largest}. These aggregations can also deter or kill intruders by forming temporary defensive aggregations, such as eastern honey bees and western honey bees killing wasps and foreign queens, respectively, by covering the intruders in a three-dimensional ball. Balls of honey bees kill an intruding wasp by contracting their flight muscles to generate heat, raising the temperature around the wasp; the mechanism by which western honey bees kill foreign queens is unknown \cite{ono1995unusual, gilley2001behavior}. 

Other aggregations temporarily form to help the individuals survive in and navigate their environment, such as clusters of ladybird beetles, feeding piles of maggots, and entangled worms. An important function of these aggregations is maintaining a comfortable internal environment for the individuals, including temperature and humidity levels \cite{copp1983temperature, rivers2011physiological, heaton2014quantifying, aydin2020collective}. This type of aggregation can help the colony survive and reach a more favorable environment when it is displaced from its home: fire ants aggregate into thin waterproof rafts on water surfaces or towers on vegetation to survive floods \cite{foster2014fire, phonekeo2017fire}, and, similarly, western honey bees aggregate into three-dimensional hanging swarms while searching for a more permanent nest site \cite{seeley2010honeybee, peleg2018collective}. Some aggregations take advantage of other organisms to survive. For instance, Mojave Desert blister beetle larvae clump on blades of grass to mimic a female bee and parasitize male bees, which carry them to the food stores of real female bees \cite{saul2006phoretic, saul2018deceptive, hafernik2000beetle}. Aggregations can also travel to more favorable environments by walking or crawling, such as the processions of larvae \cite{brues1951migrating,white2020one,lashley2018indirect}. Finally, a small fraction of invertebrates aggregating can help the entire colony travel over rough terrain, such as marching army ants that link their bodies into bridges across gaps to shorten the path of the other ants \cite{reid2015army,graham2017optimal}.

Some invertebrate aggregations, such as fire ant rafts and entangled worm blobs, have recently been investigated from a soft matter standpoint, but many unanswered questions about their behavior remain \cite{mlot2011fire,aydin2020collective,deblais2020rheology}. Other invertebrate aggregations, such as piles of maggots \cite{rivers2011physiological, heaton2014quantifying} or crawling ``mega-larvae'' made up of fungus gnat larvae \cite{jones1893rope, williston1894explanation, brues1951migrating}, have been investigated from an entomology viewpoint, but the local behavioral rules and their underlying physics are not well understood. To provide inspiration for studying these aggregations, we will now turn our attention to the existing physics-based investigations, currently available for a limited group of invertebrate aggregations.

\section{Physical properties of individuals}

To understand the underlying behavioral physics of an aggregation of invertebrates, we must first understand its constituting individuals, as their physical properties and interactions with their local environment and each other determine the behavior of the aggregation.

\subsection{The geometry of an individual}

\begin{figure}[!htb]
	\begin{center}
		\includegraphics[width = 6.5in,keepaspectratio=true]
		{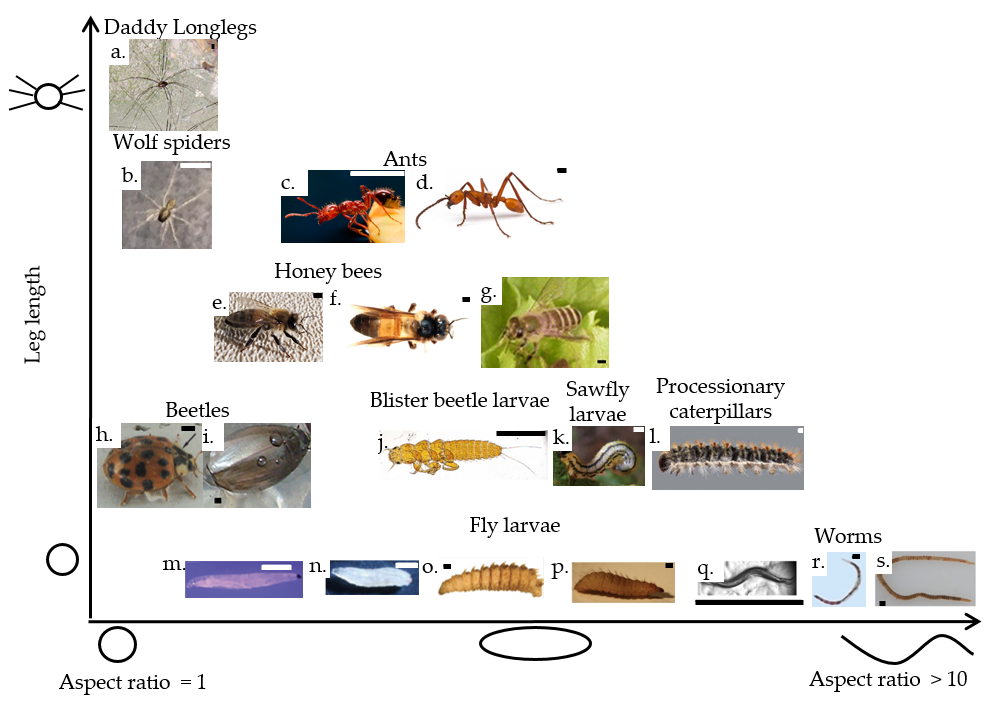}
	\end{center}  

	\caption{Invertebrates known to aggregate, organized by the aspect ratio of the individuals on the x-axis (from individuals that are rounded, like atoms or molecules, to individuals that are long and flexible, like polymers (\ref{g:polymer})) and their leg length on the y-axis (from individuals with no legs to individuals with very long legs). The scale bar in each image is 1 mm long. 	
	a. Daddy longlegs (Opiliones: \textit{Leiobunum}) \cite{wijnhoven2011notes}.
	b. Juvenile wolf spider (\textit{Pardosa saltans}) \cite{laino2020residual}.
	c. Fire ant (\textit{Solenopsis invicta}). Photograph by Tim Nowack and David L. Hu.
	d. Army ant (\textit{Eciton burchellii}). Photograph by Daniel Kronauer.
	e. Western honey bee (\textit{Apis mellifera}).
	f. Giant honey bee (\textit{Apis dorsata}) \cite{kitnya2020geographical}. 
	g. Japanese honey bee (\textit{Apis cerana japonica}) \cite{yokoi2015visitation}. 
	h. Ladybird beetle (\textit{Harmonia axyridis}) \cite{goetz2008harmonia}.
	i. Whirligig beetle (Gyrinidae: \textit{Dineutus sublineatus}) \cite{lin2012visual}.
	j. Blister beetle larva (Coleoptera: \textit{Meloe}) \cite{topitzhofer2018first}.
	k. Spitfire sawfly larva (Symphyta: \textit{Hemichroa crocea}) \cite{boeve2015multimodal}.
	l. Pine processionary caterpillar (\textit{Thaumetopoea pityocampa}) \cite{battisti2015natural}.
	m. Darkwinged fungus gnat larva (Sciaridae: \textit{Bradysia odoriphaga}) \cite{shi2017control}.
	n. Fruit fly larva (\textit{Drosophila melanogaster}) \cite{agianian2007preliminary}.
	o. Blow fly larva (\textit{Calliphoridae albiceps}) \cite{szpila2009key}.
	p. Black soldier fly larva (\textit{Hermetia illucens}) \cite{shishkov2019black_1}.
	q. \textit{Caenorhabditis elegans} worm \cite{corsi2015transparent}.
	r. Sludge worm (\textit{Tubifex tubifex}) \cite{deblais2020rheology}. 
	s. Blackworm (\textit{Lumbriculus variegatus}). Photograph by Yasemin Ozkan Aydin and Saad Bhamla.
	}
	\label{individual}
\end{figure}

The shape and size of aggregating invertebrates affect how they link together to structure their aggregation. We highlight the range of geometries of the nineteen invertebrate species under consideration in \fig{individual}, organized by their aspect ratio (\ref{g:aspect}) on the x-axis and the length of the individual's legs relative to body length on the y-axis. 


Investigations of granular mechanics (\ref{g:granular}) and liquid crystals (\ref{g:LC}) reveal that densely packed elongated particles tend to align with one another. How they align depends on particle geometry, activity, and packing density. Particles with low aspect ratios or a low packing density result in an isotropic, or disordered, packing structure (\ref{g:isotropic}). When the packing density of the particles reaches a threshold, particles with higher aspect ratios align with one another in nematic or smectic phases (\ref{g:nematic}, \ref{g:smectic}) and can even form a helical structure in a cholesteric phase (\ref{g:cholesteric}) \cite{ramaswamy2010mechanics, marchetti2013hydrodynamics, bolhuis1997tracing}.

Some arrangements of invertebrates in aggregations are analogous to liquid crystal phases, with individual aspect ratio as a key parameter. The ladybird beetle (\textit{Harmonia axyridis}) and whirligig beetle (Gyrinidae) shown in \fig{individual}{\bf(h-i)} are round with an aspect ratio of less than 2, and their aggregations are likely to be isotropic. On the other hand, blister beetle larvae (\textit{Meloe} sp.), sawfly larvae (Symphyta), and processionary caterpillars (\textit{Thaumetopoea pityocampa} and other Lepidoptera caterpillars), as in \fig{individual}{\bf(j-l)}, and fly larvae (Diptera), as in \fig{individual}{\bf(m-p)}, are longer, with an aspect ratio close to 5, and thus are more likely to align with one another into a nematic phase.

If particles are long and flexible, they can wrap around one another and entangle (\ref{g:entanglement}), as is often seen with polymers (\ref{g:polymer}) \cite{watanabe1999viscoelasticity}. For a living example of entangled polymers, we can consider the worms in \fig{individual}{\bf(q-s)} (\textit{Caenorhabditis elegans, Tubifex tubifex, Lumbriculus variegatus}). These worms are long, with an aspect ratio greater than 10, and also flexible as seen in the U-shape of the sludge worm and blackworm in \fig{individual}{\bf(r-s)}. Instead of aligning in aggregations, these worms can wrap their bodies around each other to entangle.

In addition to the body aspect ratio, the leg length of individuals (and whether they have legs at all) can impact the resulting aggregation. Invertebrates that do not have legs, such as fly larvae and worms in \fig{individual}{\bf(m-s)}, or that have short legs, such as the beetles in \fig{individual}{\bf(h-j)}, sawfly larvae in \fig{individual}{\bf(k)} and caterpillars in \fig{individual}{\bf(l)} align or entangle with one another to aggregate. In contrast, the connections between invertebrates that can grip each other with claws or adhesive pads on their legs \cite{foster2014fire}, such as the fire ants (\textit{Solenopsis invicta)} and \textit{Eciton} army ants (\textit{Eciton burchellii}) in \fig{individual}{\bf(c-d)}, and  western honey bees, Giant honey bees, and Japanese honey bees (\textit{Apis mellifera, Apis dorsata, Apis cerana japonica}) in \fig{individual}{\bf(e-g)}, are not limited to the geometry and flexibility of their bodies. Strong leg--leg connections, such as bonds between fire ants that can bear 400 ant weights, can give more structure to an aggregation than merely the friction of animals on one another in a pile \cite{mlot2011fire}. Meanwhile, arthropods such as daddy longlegs (Opiliones) and juvenile wolf spiders (Lycosidae) as in \fig{individual}{\bf(a-b)} resemble a star polymer (\ref{g:star}) \cite{ren2016star} with their relatively long legs; how these legs affect the structure of their aggregation is not well understood.

\clearpage

\subsection{The behavior of an individual}
The phases of alignment between particles in an inactive granular material (\ref{g:inactive}, \ref{g:granular}) depend on the temperature of the particles as well as the packing density and aspect ratio. In general, increasing the temperature (and, by proxy, the random motion of the particles), tends to decrease the order in the system (\ref{g:order}) and change the phase of the system from nematic and smectic phases to isotropic \cite{bolhuis1997tracing, cladis1975new}. Adding directed activity by giving the particles an internal source of energy, propelling them in some direction (\ref{g:active}), adds new options for ordered phases. The options include particles that only move in one direction aligning along their direction of motion in a polar state (\ref{g:polar}, \ref{g:polarorder}) or particles that can move forward and backward, forming a nematic state due to their activity (\ref{g:apolar}) \cite{marchetti2013hydrodynamics}. The activity of invertebrates can cause them to arrange into these states; however, their motion can also deviate from self-propulsion in one direction, leading to even more possible states.

The interactions of invertebrates with one another and with their environment determine how they use their physical properties to aggregate. In species with complex social dynamics, the caste of the individuals affects whether an aggregation forms. For example, honey bee swarms are made up of a queen and several thousand workers \cite{peleg2018collective}, and fire ant rafts are easier to create if eggs are not included \cite{mlot2011fire}. In non-eusocial insects such as fly larvae or sawfly larvae, age or larval stage might similarly affect whether individuals aggregate \cite{shishkov2019black_1,fletcher2007vibrational}.

How individuals respond to their environment can induce or suppress aggregation. Sometimes, the same stimulus can result in the formation of one type of invertebrate aggregation, and the dissolution of another. 

Many invertebrates prefer dimly lit environments, but the reactions of different invertebrate species to light are vastly different. Bright light disrupts the foraging and feeding aggregations of fruit fly, sawfly, and black soldier fly larvae, leading them to individually seek shelter. Red or infrared light, which fly larvae, sawfly larvae, and many other insects are less sensitive to, is necessary to observe these aggregations \cite{keene2012seeing, fletcher2007vibrational, shishkov2019black_1}. In contrast, blackworms entangle into clumps when their environment is brightly lit to hide individuals inside the mass of worms \cite{aydin2020collective}.

The aggregations of aquatic invertebrates are influenced by water chemistry. For instance, decreased oxygen concentration slows down \textit{C. elegans} individuals, causing them to aggregate \cite{sugi2019c, demir2020dynamics}. On the other hand, sludge worm aggregations help limit the exposure of the worms inside to toxic dissolved oxygen in the water \cite{deblais2020phase}. Other chemicals added to the water are absorbed through the skin of the invertebrates in it and can change their behavior. For example, adding alcohol to water containing sludge worms results in decreased worm activity \cite{deblais2020rheology}. Finally, aggregating can help aquatic individuals survive in dry environments, such as blackworms that entangle to prevent desiccation \cite{aydin2020collective}.

Food is another powerful motivator for aggregation. The presence of food can cause fly larvae and other invertebrates to aggregate while feeding, while the lack of food in the environment sometimes results in foraging aggregations, such as caterpillar processions, traveling piles of maggots, sawfly larvae, or fungus gnat larvae, and army ant bridges \cite{jones1893rope, williston1894explanation, brues1951migrating, sutou2011recent,uemura2020movement, white2020one,fitzgerald2003role, shishkov2019black_1,reid2015army}.

Finally, invertebrates are primarily ectothermic, and aggregating can help them maintain a comfortable temperature. There are many documented examples of aggregations of bees, ants, ladybird beetles, daddy longlegs, and wolf spiders thermoregulating for the comfort of the individuals inside \cite{copp1983temperature, machado2002alarm, coddington1990mass, franks1989thermoregulation, cully2004self,peters2021thermoregulatory, baudier2019plastic}.

These environmental factors cause invertebrates with similar aspect ratio and leg type to create vastly different aggregations, which we discuss in Sections 4 and 5.

\section{Physical properties of aggregations}

\begin{SCfigure}
		\includegraphics[width = 4in,keepaspectratio=true]
		{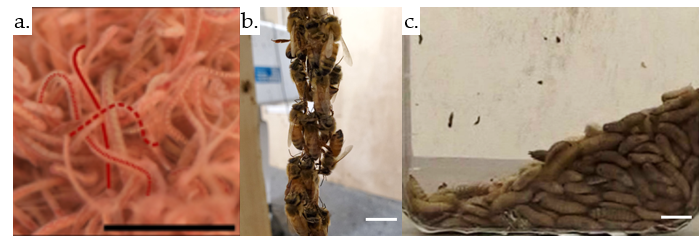}
	\caption{ (a) Close up of entangled sludge worm blob \cite{deblais2020rheology} (b). A chain of bees linked to each other with their legs (c). Side view of a pile of black soldier fly larvae against the side of a container \cite{shishkov2020synchronizing}. The scale bar in each panel is 1 cm long.}
	\label{bonds}
\end{SCfigure}

\begin{figure}[!htb]
	\begin{center}
		\includegraphics[width = 5in,keepaspectratio=true]
		{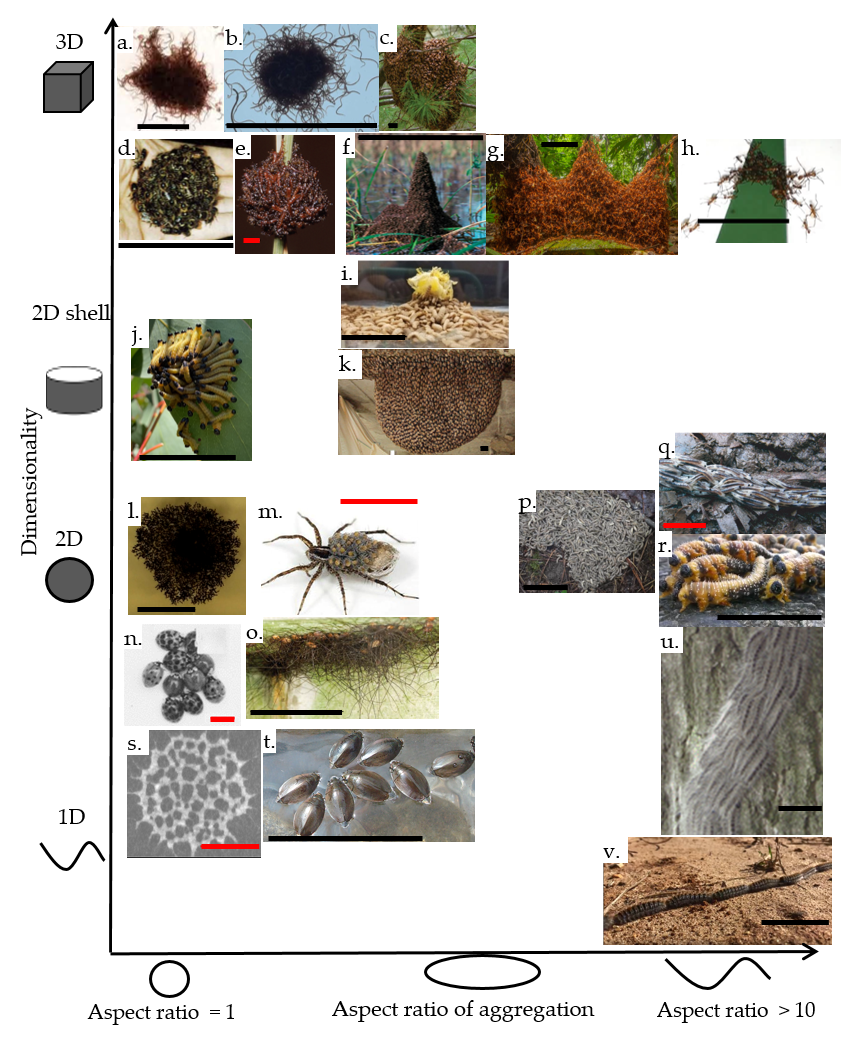}
	\end{center}  
	\vspace{-4ex}

	\caption{Characteristics of aggregations of invertebrates. We organize aggregations by aspect ratio on the x-axis and dimensionality (from a one-dimensional linear procession to three-dimensional balls and clusters) on the y-axis. Black scale bars are 5 cm long, red scale bars are 0.5 cm long. 
  a. Blackworm (\textit{Lumbriculus variegatus}) blob \cite{aydin2020collective}. For video, see {\bf Supplementary Video \ref{v:worm}}.
	b. Sludge worm (\textit{Tubifex tubifex}) blob \cite{deblais2020phase}.
	c. Western honey bee swarm. (\textit{Apis mellifera}) \cite{peleg2018collective}.
	d. Japanese honey bee (\textit{Apis cerana japonica}) heat ball \cite{ono1995unusual}. 
	e. Blister beetle larvae (\textit{Meloe franciscanus}) clumped on a stem \cite{saul2006phoretic}. 
	f. Fire ant (\textit{Solenopsis invicta}) tower \cite{phonekeo2017fire}.
	g. Army ant (\textit{Eciton burchellii}) bivouac. Photograph by Daniel Kronauer.
	h. Army ant (\textit{Eciton burchellii}) bridge \cite{reid2015army}. For video, see {\bf Supplementary Video \ref{v:antbridge}}.
	i. Black soldier fly larva (\textit{Hermetia illucens}) pile \cite{shishkov2019black_1}.
	j. Sawfly larva (Symphyta) aggregation. Frame from video by David Yeates. For video, see {\bf Supplementary Video \ref{v:sawflytwitch}}. 
	k. Giant honey bee (\textit{Apis dorsata}) swarm \cite{kastberger2014giant}. For video, see {\bf Supplementary Video \ref{v:giantbee}}. 
	l. Fire ant (\textit{Solenopsis invicta}) raft \cite{mlot2011fire} For video, see {\bf Supplementary Video \ref{v:antraft}}. 
	m. Juvenile wolf spiders (Lycosidae: \textit{Pardosa saltans}) on their mother's back \cite{trabalon2018embryonic}.
	n. Ladybird beetle (\textit{Harmonia axyridis}) swarm \cite{nalepa2007infection}.
	o. Daddy longlegs (Opiliones: \textit{Leiobunum}) aggregation \cite{shear2009opiliones}. 
	p. Blow fly larvae (Calliphoridae)) migration \cite{lashley2018indirect}. For video, see {\bf Supplementary Video \ref{v:blowfly}}. 
	q. Darkwinged fungus gnat larva (Sciaridae) aggregation \cite{sutou2011recent}. For video, see {\bf Supplementary Video \ref{v:fungusgnat}}. 
  r. Sawfly larvae in a migrating pile. Frame from video by Ben R. Fitzpatrick. For video, see {\bf Supplementary Video \ref{v:sawflymig}}. 
	s. Wild type \textit{C. elegans} aggregation on agar \cite{sugi2019c}. 
	t. Whirligig beetle (Gyrinidae: \textit{Dineutus sublineatus}) aggregation \cite{lin2012visual}. 
	u. Oak processionary caterpillar (\textit{Thaumetopoea processionea}) aggregation \cite{maronna2008lepidopterism}. 
	v. Procession of pine processionary caterpillars (\textit{Thaumetopoea pityocampa}) \cite{uemura2020movement}. For video, see {\bf Supplementary Videos \ref{v:onecaterpillar}} and {\bf \ref{v:twocaterpillars}}. 
	}
	\label{aggregation}
\end{figure}

The physical and behavioral differences in the invertebrates described in Section 2 result in a diverse variety of aggregations. In this section, we first review aggregations by how the individuals are connected in \fig{bonds}, from entanglement to surface contact. We then review aggregation geometry by the aspect ratio and dimensionality in \fig{aggregation}, similarly to the organization of individuals by aspect ratio and leg length. These categorizations help understand the behavior of the dense invertebrate aggregations presented in \fig{individual} as materials.

\clearpage

\subsection{How Bond Types Affect Aggregations}

The internal structure of an aggregation depends on how the individuals inside are connected. We categorize the active bonds in an aggregation into three main types, shown in \fig{bonds}: entanglement, strong leg--leg bonds, and friction, although some aggregations may have properties that fall into more than one category.

In entangled (\ref{g:entanglement}) aggregations, long flexible individuals wrap their bodies around one another, such as the worms in \fig{bonds}{\bf(a)}. These aquatic worms entangle into a roughly spherical ``blob'', shown in \fig{aggregation}{\bf(a--b)} and {\bf Supplementary Video \ref{v:worm}} \cite{deblais2020phase,aydin2020collective}. The bodies of smooth individuals would slide on one another, causing the individuals to disentangle and drift apart. Sludge worms have small bristles on their bodies that prevent sliding and reinforce the entangled aggregation \cite{deblais2020phase}.

In aggregations connected by strong, reversible leg--leg bonds, individuals use claws and adhesive pads on their legs to grip one another, such as the bees in \fig{bonds}{\bf(b)}. These bonds give individuals finer control of their position and allow them to modify the aggregation by breaking and reforming bonds in response to local stimuli. These bonds are critical to the formation of these aggregations: deactivating fire ant tarsal pads by coating ants in baby powder reveals that ants require tarsal pad connections to link together \cite{foster2014fire, phonekeo2017fire}. Examples of invertebrates aggregating using leg--leg bonds include army ants in bridges and bivouacs, fire ants in rafts or towers, western honey bees in swarms, giant honey bees covering the surface of their nests, and blister beetle larvae on stalks of grass, as shown in \fig{aggregation}{\bf(c--f, i)} \cite{mlot2011fire, peleg2018collective,kastberger2008social,hafernik2000beetle,reid2015army}.

In aggregations with frictional bonds, individuals touch one another but do not form strong bonds with one another, such as the black soldier fly larvae in  \fig{bonds}{\bf(c)}. These individuals may either not be able to grab one another with their legs or have no legs at all, and are not long and flexible enough to entangle their bodies with one another. This can be thought of as having a short-range attraction potential \ref{g:attract}. The result is either individuals piled up on one another or clustered on a surface and touching each other with their skin. These piles can be three-dimensional, such as feeding fly larvae, daddy longlegs and beetles clumping for warmth, spiderlings on their mother's back, and sawfly or fungus gnat larvae piling over one another in processions, in \fig{aggregation}{\bf (i,m--o,p--r)} \cite{rivers2011physiological, shishkov2019black_1, lashley2018indirect, jones1893rope, williston1894explanation, brues1951migrating, copp1983temperature, machado2002alarm,coddington1990mass,brach1976subsocial}. In the two-dimensional limit of a pile, individuals cluster next to each other on a surface. This type of aggregation is found in sawfly larvae covering leaves, whirligig beetles on water surfaces, and processionary caterpillars, as in \fig{aggregation}{\bf(j,t,u,v)} \cite{lin2012visual, voise2011capillary,uemura2020movement, white2020one,fitzgerald2003role,boeve2015multimodal, fletcher2007vibrational}.

In addition to these active bonds, individuals use the physical forces between themselves and their surroundings to aggregate, such as the surface tension on the water-air interface (\ref{g:surface}). Networks of wild type \textit{C. elegans} worms, as in \fig{aggregation}{\bf(s)} form primarily through the interaction of physical forces: the aggregation forms when individual motion is reoriented along the same axis by collisions, and the resulting networks are held together by surface tension \cite{sugi2019c, demir2020dynamics}. Other aggregations form with the assistance of mechanical forces, but use their activity to fine-tune the aggregation structure: for instance, whirligig beetles are initially drawn together by capillary forces and move within the aggregation to find preferred positions for foraging or conserving energy \cite{lin2012visual, voise2011capillary, romey2008optimal}. Finally, individuals can intentionally aggregate to change the force distribution on their bodies. Rafting fire ants trap a bubble of air within the raft, increasing the hydrophobicity and buoyancy of the raft. This allows the raft to float despite individual ants being too large to be supported by surface tension \cite{mlot2011fire}. To understand how an invertebrate aggregation is structured, it is necessary to understand the combination of these passive bonds and the entangled, leg--leg, and frictional active bonds between individuals.

Now that we categorized the type of bonds that exist within an aggregation, we can move on to categorizing the aggregation's shape in Section 4.2.

\subsection{The geometry of an aggregation}

Here we highlight twenty-two sample dense invertebrate aggregations and organize them by their geometry in \fig{aggregation}, with the aspect ratio of the aggregation (\ref{g:aspect}) on the x-axis, and dimensionality on the y-axis (aggregations where the individuals span an axisymmetric volume are ``three-dimensional''). The aspect ratio arises from the aggregation's function and the surroundings, and varies from the rounded blobs in \fig{aggregation}{\bf(a-b)} to the long chains of processionary caterpillars in \fig{aggregation}{\bf(v)}. Rounded aggregations, such as the bee swarms  and army ant bivouacs shown in \fig{aggregation}{\bf(c,d,f,g,l)}, reduce surface area. Other aggregations, such as ant bridges and moving piles of larvae in \fig{aggregation}{\bf(h,p-v)} are elongated as the organisms collectively travel from place to place \cite{reid2015army,brues1951migrating,white2020one}. 

The dimensionality of an aggregation describes the space that an aggregation occupies, from aggregations occupying volumes in \fig{aggregation}{\bf(a-h)}, to aggregations several individuals thick on flat surfaces or shells in\fig{aggregation}{\bf(i-u)}, to the one-dimensional chain of caterpillars in \fig{aggregation}{\bf(v)}. Three-dimensional aggregations enclose a large number of individuals in their internal environment and can hide individuals from predation, such as the worker bees surrounding the queen in a honey bee swarm or army ants protecting the queen and brood in a bivouac \cite{franks1989thermoregulation, heinrich1981mechanisms, kronauer2020army, cully2004self}. Aggregations spread out on exposed surfaces have more active ways of protecting the individuals inside, such as the tail flicking of sawfly larvae and giant honey bee shimmering in \fig{aggregation}{\bf(j-k)} \cite{kastberger2008social,fletcher2007vibrational}. 

Aggregations can change their shape and dimensionality to adapt to their environment: for instance, fire ants flatten their rafts to take advantage of surface tension to trap an air bubble with the hydrophobic ant cuticle and stay afloat \cite{mlot2011fire}. Next, we discuss the properties of aggregations resulting from their bonds and shape and how they change these properties in Section 5.

\section{Aggregations as active materials}

In this section, we describe how physics-based approaches are used to understand the behavior of dense invertebrate aggregations. Some of these approaches take a top-down view of the aggregation to learn its bulk properties - for example, studying the viscoelasticity of an aggregation with rheology or the flow of individuals within it with particle image velocimetry \ref{g:PIV}. Other studies take a bottom-up view, tracking the behavior of individuals to understand how the aggregation functions as a whole. We first discuss the analogy between properties and states of soft materials to invertebrate aggregations in Section 5.1 and then how individuals move in an aggregation in Section 5.2.

\subsection{Material properties of aggregations}

Many invertebrate aggregations can be thought of as self-assembled living materials \cite{vernerey2018fire, tennenbaum2016mechanics, aydin2020collective,tennenbaum2017activity}, and the tools for understanding materials can be useful for understanding these aggregations. Are these invertebrate aggregations liquid or solid? In many cases, they are both. Treating very dense aggregations, in which invertebrates are tightly packed with one another, as viscoelastic materials \ref{g:viscoelasticity} can elucidate their solid (elastic, \ref{g:elasticity}) or liquid (viscous, \ref{g:viscosity}) properties. 

Rheology (\ref{g:rheology}) shows that fire ant and sludge worm aggregations have a viscoelastic response to shear. Very dense fire ant aggregations are primarily elastic; less crowded fire ant aggregations are both elastic and viscous and exhibit shear-thinning (their viscosity decreases with higher shear strain, \ref{g:thin}) at high deformation rates \cite{tennenbaum2016mechanics, tennenbaum2017activity}. Sludge worm blobs can be thought of as entangled active polymers (\ref{g:polymer}, \ref{g:entanglement}). Like inactive entangled polymers, these blobs are shear-thinning \cite{deblais2020rheology}. Similarly, constant strain (\ref{g:strain}) compression experiments of black soldier fly larvae reveal how the aggregation reacts to external forces from material, such as good scraps or compost, piled on them. When compressed, inactive black soldier fly larvae have viscoelastic properties; meanwhile, active larvae respond to applied forces in seconds to alleviate the compressive forces on individual larvae \cite{shishkov2019black_2}. 

The effects of activity on  viscosity and shear-thinning are different for fire ant and sludge worm aggregations. Fire ant aggregations have a higher viscosity than sludge worm blobs: the viscosity of fire ants decreases from $10^6$ to $10^0 $ Pa s as the shear rate increases from $10^{-4}$ to $10^2 $ 1/s, while the viscosity of sludge worms decreases from $10^2$ to $10^{-1}$ Pa s as the shear rate increases from $10^{-3}$ to $10^1$ 1/s \cite{tennenbaum2016mechanics, deblais2020rheology}. Shear-thinning of fire ant aggregations at high deformation rates is thought to happen as ants break their leg--leg bonds to avoid being damaged \cite{vernerey2018fire}. Meanwhile, sludge worm activity is inversely correlated with viscosity at low shear rates, most likely because the activity rearranges and disentangles the worms. At high shear rates, increased worm activity is positively correlated with viscosity. Thus, sludge worm shear thinning is effectively decreased by activity \cite{deblais2020rheology}.

These experiments with ants, worms, and fly larvae show that aggregations have material properties that can be measured by applying external forces. They also demonstrate that individual activity has different effects on the properties of different systems. Insights gained from these studies both explain how aggregated invertebrates respond to the forces they encounter in nature, and how adding activity would affect the properties of a non-living granular material.

Some nvertebrate aggregations  control their temperature for the comfort of the individuals inside instead of passively heating or cooling with their environment like  an inactive material, so the discussion of their thermal properties goes beyond measurements of properties such as heat conductance. Thermal imaging and arrays of temperature sensors are powerful tools to investigate invertebrate thermoregulation. Some aggregations thermoregulate to raise the temperature to a certain threshold. For instance, eastern honey bee species including \textit{Apis cerana japonica} and \textit{Apis dorsata} can kill a wasp by covering it in a three-dimensional ``heat ball'', as in \fig{aggregation}{\bf(d)}. They raise the temperature of the wasp to 47$^\circ$ C \cite{ono1995unusual, kastberger2008social}, but no higher, since the lethal temperature of a honey bee is 48 - 50$^\circ$ C \cite{ono1995unusual}. Fly larva piles generate heat to increase larval metabolism \cite{johnson2014infrared, johnson2014tracking, tomberlin2009development} up to a lethal temperature threshold (which depends on the species) \cite{rivers2011physiological}. Other aggregations, such as army ant bivouacs and  western honey bee swarms, aim to keep the temperature in the middle of a comfortable range than at the highest end of it. Army ants in bivouacs use their metabolism to generate heat and open air channels inside the bivouac to cool it. Similarly,  western honey bee swarms keep a constant temperature inside the swarm by contracting the swarm in cold weather and expanding it in the heat \cite{peters2021thermoregulatory, heinrich1981mechanisms, cully2004self}. Army ant bivouac and honey bee swarm changes in response to ambient conditions are not directly correlated to ambient temperature. Army ant bivouacs only raise the internal temperature if necessary for the survival of the brood inside and otherwise lower their metabolic rate to survive at a colder temperature \cite{baudier2019plastic}, and honey bee swarms respond faster and maintain a more consistent swarm shape when the environment is cooled than heated \cite{peters2021thermoregulatory}. This likely helps the ants or bees conserve energy and account for mechanical constraints on their structure while maintaining acceptable internal conditions for the constituent individuals.

So far, we have reviewed invertebrate aggregations through the lens of soft materials. However, these aggregations are made up of living, sentient individuals who can initiate internal motion and collective locomotion that goes beyond what conventional soft materials can do.

\subsection{Motion of an aggregation}

\begin{figure}[!htb]
	\begin{center}
		\includegraphics[width = 6.5in,keepaspectratio=true]
		{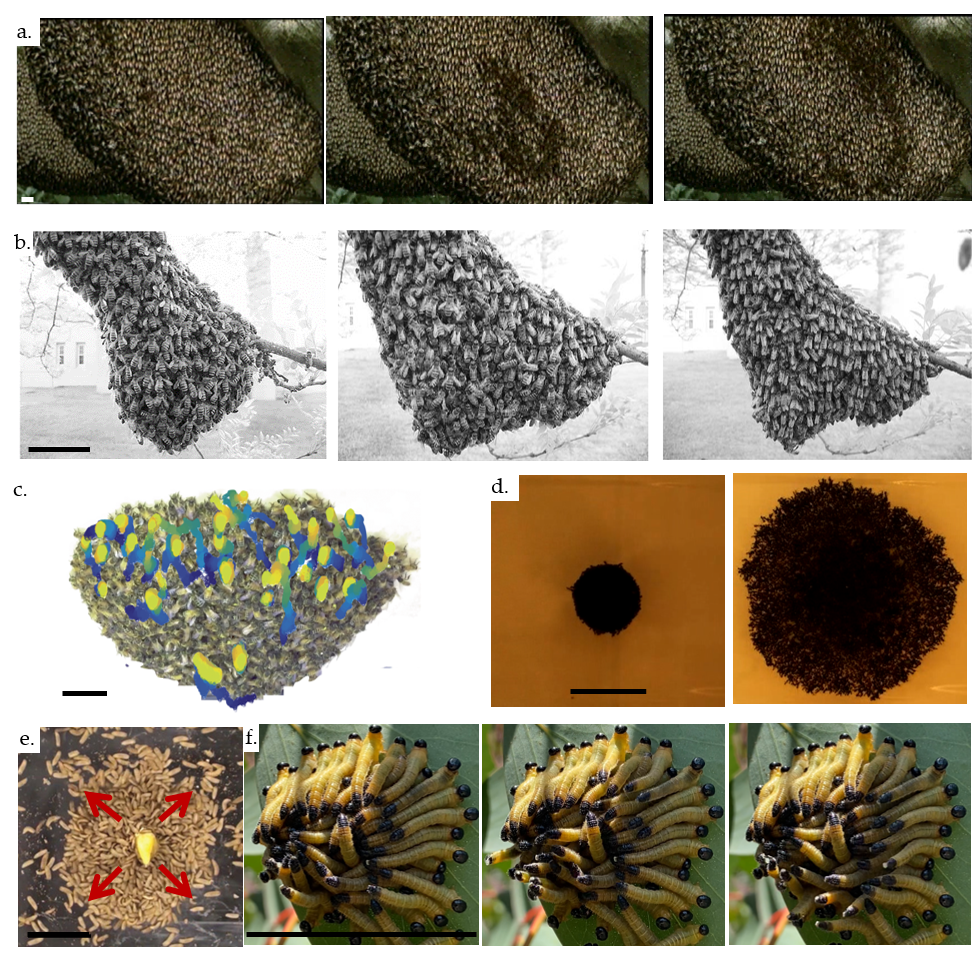}
	\end{center}  
	\caption{
  Examples of individual motion within invertebrate aggregations. All scale bars are 5 cm long.
	a. Sequence of frames showing giant honey bees ``shimmering''. Left: bees on the surface of their nest. Center: bees detect a threat and begin shimmering. Right: the shimmering wave propagates through the bees \cite{kastberger2008social}. See {\bf Supplementary Video \ref{v:giantbee}} for video.
	b. Western honey bees reconfigure their swarm depending on the weather. Left: bees are closely packed on a cold, dry morning. Center: bees expand their swarm in the warmer afternoon. Right: bees rearrange into a protective curtain from rain \cite{cully2004self}. 
	c. Trajectories of western honey bees on the surface of a swarm under mechanical shaking show bees moving up to reinforce the swarm. Tracks are colored by time, with blue at the start of the experiment and yellow at the end. Image from Ref. \cite{peleg2018collective}.
	d. Fire ants change the shape of their raft. Left: A ball of ants is placed on the water surface. Right: Ant raft flattens into a thin pancake. See {\bf Supplementary Video \ref{v:antraft}} for full video \cite{mlot2011fire}.
	e. Top view of a pile of black soldier fly larvae eating an orange slice. Larvae fall down the top of the pile as indicated by arrows. Fresh larvae crawl towards the food on the floor to replace those falling \cite{shishkov2019black_1}.
	f. Sawfly larvae flick their posterior ends, or ``tails'', in an alarm display. Left: sawflies are static before a flick. Center: Sawflies on the leftmost edge of the leaf begin flicking their tails. Right: sawflies on the right edge of the leaf flick their tails. Frames from video by David Yeates. See {\bf Supplementary Video \ref{v:sawflytwitch}} for video.
	}
	\label{internal}
\end{figure}

\begin{figure}[!htb]
	\begin{center}
		\includegraphics[width = 6.5in,keepaspectratio=true]
		{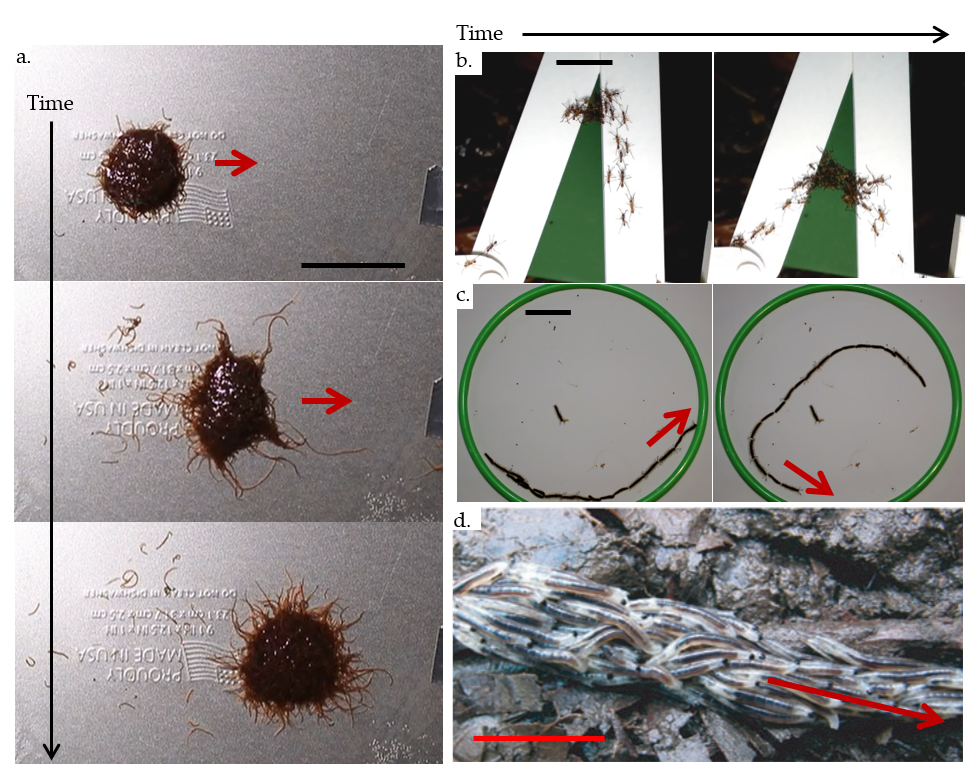}
	\end{center} 
	\caption{
 Examples of aggregations in which the collective actions of the individuals move the center of mass. Scale bars in (a-c) are 5 cm long; scale bar in (d) is 0.5 cm long.
  a. Blackworm blob moves towards cold water. Top: The worm blob is subject to a temperature gradient at the start of the experiment. Center: the blob moves towards the colder side of the water on the right. Bottom: the warm blob reaches the cold water. Images from Ref. \cite{aydin2020collective}. Full video is shown in {\bf Supplementary Video \ref{v:worm}}.
  b. An army ant bridge forms and grows to span a wider gap in the path as more ants join \cite{reid2015army}. Full video is shown in {\bf Supplementary Video \ref{v:antbridge}}. 
  c. Time-lapse of a trail of two species of processionary caterpillars in a procession. Full video is shown in {\bf Supplementary Video \ref{v:twocaterpillars}}. Left and right frames show the procession at different times. Video by Stefanie White and Amy Deacon.
	d. A snapshot of fungus gnat larvae moving in the direction of the arrow \cite{sutou2011recent}. A video of migrating fungus gnats is shown in {\bf Supplementary Video \ref{v:fungusgnat}}. 	
	}
	\label{COM}
\end{figure}

Individuals within an aggregation are often far from static, distinguishing these aggregations from piles of grains. We consider two types of motion: first, internal motion that can change the aggregation surface or bulk structure  (\fig{internal}), and, second, motion that propels the entire aggregation forward  (\fig{COM}). We present the known time scales and speeds of motion of aggregations in {\bf Supplementary Table \ref{veltable}}.

We first consider internal motion within an aggregation in  (\fig{internal}). Some invertebrates move in fast, intermittent bursts lasting seconds to deter intruders, and then return to their original state. For example, giant honey bee colonies (\fig{aggregation}{\bf(k)}) keep hornets away by ``shimmering'' (\fig{internal}{\bf(a)} and {\bf Supplementary Video \ref{v:giantbee}}). When quiescent bees detect a hornet, several bees initiate shimmering by flipping their abdomens upwards, as in the center image. The wave of flipping abdomens travels around the nest \cite{kastberger2008social}. This wave is initiated by specialized groups of bees, and other bees join by following their neighbors \cite{schmelzer2009special, kastberger2012join, kastberger2014speeding}. Shimmering vibrates the nest, which might help bees communicate about their defensive state \cite{kastberger2013social}. Similarly, sawfly larvae gathered in a clump on a leaf (\fig{aggregation}{\bf(j)}) flick their tails in synchronized bursts to ward off predators ({\bf Supplementary Video \ref{v:sawflytwitch}} and \fig{internal}{\bf(f)}). This motion is also synchronized through substrate vibrations \cite{boeve2015multimodal, fletcher2007vibrational}. Finally, aggregating whirligig beetles (\fig{aggregation}{\bf(t)}) suddenly swim away from each other when they sense danger, and take some time to regroup (termed a ``flash expansion'') \cite{romey2015flash}. 

In some invertebrate aggregations, slower transient motion (on the order of minutes to hours) can change the shape and density of an existing aggregation or form a new aggregation. For an example of shape change, western honey bee swarms (\fig{aggregation}{\bf(c)}) maintain a comfortable bulk temperature by changing the density of the swarm. The time series in \fig{internal}{\bf(b)} shows that bees in a swarm are tightly packed in cold weather and loosely packed in warm weather \cite{cully2004self, heinrich1981mechanisms, peters2021thermoregulatory}. When a gust of wind shakes the swarm, bees move up to its base where the strain is highest to reinforce it, flattening the cluster until the shaking stops (\fig{internal}{\bf(c)}) \cite{peleg2018collective}. Similarly to bee swarms, rafting fire ants keep the raft floating and cohesive in changing water conditions by breaking and forming new connections \cite{mlot2011fire, mlot2012dynamics}. These fire ant rafts spread from a ball into a flat, water-repellent raft in minutes and float for days until they reach land (\fig{aggregation}{\bf(l)}, \fig{internal}{\bf(d)} and {\bf Supplementary Video \ref{v:antraft}}). Ants walk towards the raft edges and attach at the water surface, forming a ``treadmill'' to expand the raft outwards \cite{mlot2011fire, wagner2021treadmilling}. These aggregations can be thought of as living adaptive materials that respond to their environment \cite{walther2020responsive}.

Analogies between the formation of aggregations of invertebrates to phase separation (\ref{g:phase}) of non-living particles reveal unique properties of living aggregations. Wild-type \textit{C. elegans} worms align through collisions with one another to aggregate into a nematic phase (\ref{g:nematic}), as in \fig{aggregation}{\bf(s)} \cite{sugi2019c, demir2020dynamics}. This process is an example of motility-induced phase separation (\ref{g:MIPS}) in a living organism \cite{cates2015motility}. The process by which sludge worms that are initially spread out in water assemble into a blob, shown in \fig{aggregation}{\bf(b)}, is similar to polymer phase separation, with many smaller blobs forming and coalescing over time. Unlike inactive polymer phase separation (\ref{g:inactive}, \ref{g:polymer}), which is caused by large polymer aggregates having a lower surface energy than small polymer aggregates (\ref{g:surface}), sludge worm phase separation is caused by small blobs actively merging to form large blobs \cite{deblais2020phase}. These examples highlight the similarities and differences between living aggregations and non-living materials. Existing theories of active matter can help explain \textit{C. elegans} aggregation behavior \cite{demir2020dynamics}; however, new models of active entanglement are necessary to describe the aggregations of worm blobs \cite{deblais2020phase}.

In other aggregations, continuous steady-state internal motion maintains the structure of the aggregation. For instance, the motion of fly larvae in a feeding pile (\fig{internal}{\bf(e)}) appears chaotic but it allows larvae to recirculate around food. Fresh individuals crawl toward food on the floor, and larvae that finished feeding fall on the top layer of the pile, as shown by treating the larvae as an active fluid using Particle Image Velocimetry (\ref{g:PIV}) \cite{shishkov2019black_1}. Similarly, individual fire ants inside towers (\fig{aggregation}{\bf(f)}) sink and are constantly replaced by ants from outside the tower as shown with x-ray videography and tracking \cite{phonekeo2017fire}. Similar internal restructuring is found in whirligig beetle aggregations, in which feeding individuals are found at the periphery of the aggregation while satiated individuals are found closer to the center \cite{romey2008optimal}.

Second, we consider the motion of the center of mass of the entire aggregation towards the individuals' desired state (\fig{COM}). In some cases, the combined velocity vectors of individuals propel an aggregation. For example, moving as an entangled aggregation helps a spherical blob of blackworms (\fig{COM}{\bf(a)} and {\bf Supplementary Video \ref{v:worm}}) follow a temperature gradient to reach their preferred cooler water. Worms sticking out of the cooler front of the blob pull it forward, while the worms on the hotter rear are coiled and reduce friction \cite{aydin2020collective}.


Some piling invertebrates travel by crawling over one another as they forage. For instance, the ``migrations'' of fungus gnat larvae (\fig{COM}{\bf(d)} and {\bf Supplementary Video \ref{v:fungusgnat}}) have been documented since the 1800s, but remain poorly understood \cite{jones1893rope, williston1894explanation, brues1951migrating, sutou2011recent}. Similar traveling piles of sawfly larvae are synchronized using body contractions and tail twitches \cite{fletcher2007vibrational}, shown in \fig{aggregation}{\bf(r)} and {\bf Supplementary Video \ref{v:sawflymig}}. Finally, blow fly larvae travel as a large disordered pile to search for a new food source (\fig{aggregation}{\bf(p)} and {\bf Supplementary Video \ref{v:blowfly}}) \cite{lashley2018indirect}. 

In the one-dimensional limit of a moving pile, processionary caterpillars walk in one-dimensional head-to-tail trails while searching for a new tree to feed from or a burrow to pupate in (\fig{aggregation}{\bf(u--v)}, \fig{COM}{\bf(c)}, and {\bf Supplementary Videos \ref{v:onecaterpillar} and \ref{v:twocaterpillars}}). These can be made up of more than one species of caterpillar and rely both on pheromones and physical touch \cite{uemura2020movement, white2020one,fitzgerald2003role}. 

A final mechanism for the motion of the center of mass can result from directed aggregation and dissipation. For example, when raiding, \textit{Eciton} army ants come across a gap in their path, a fraction of them self-organize into a bridge (\fig{COM}{\bf(b)} and {\bf Supplementary Video \ref{v:antbridge}}). The bridge becomes shorter or longer depending on the flow of ants over it, optimizing the number of raiding ants and bridge ants to maximize their foraging rate \cite{ rettenmeyer2011largest, reid2015army, graham2017optimal}.

\clearpage

\section{Discussion}
In this review, we described a number of recent advances in understanding the material properties and dynamics of invertebrate aggregations - particularly those of fire ants, army ants, western honey bees, giant honey bees, entangled worms, and fly larvae. These studies are a promising stepping stone to employing physics techniques to understand aggregations that, so far, have attracted little attention from the soft matter community. In this discussion, we describe several potential directions for furthering these studies. Drawing analogies between aggregated invertebrates and inanimate soft materials could provide a framework to conceptualize the emergent behavior of the group. 

There are a number of unanswered questions about the structure of invertebrate aggregations that can benefit from an interdisciplinary approach. It can be insightful to consider their bulk properties, such as size and packing density. The maximum size of an aggregation might be determined by physical forces, such as the maximum load that an individual in a hanging aggregation of ants or bees can support, or the maximum capillary force that can support the weight of an ant raft. Ecological factors may also play a role, such as the maximum number of individuals a colony can support with the available resources. How densely packed the individuals are within this large aggregation might be determined by the local conditions inside the aggregation. A densely packed aggregation might be more insulated against the cold, but a dense packing would be detrimental if the temperature inside is too high or the oxygen concentration is too low. 

We can also consider the bulk viscous and elastic properties of aggregations formed through entanglement or strong reversible bonds (for instance, blister beetle larva clumps) with rheology (\ref{g:rheology} ,\ref{g:viscoelasticity}). Understanding the viscoelastic properties of an aggregation shows how it withstands external forces, for instance remaining solidlike to keep the structure of the aggregation intact, or breaking the bonds between to become liquidlike and avoid damaging the individuals inside. This has proven successful with fire ant and entangled worm aggregations \cite{tennenbaum2016mechanics, deblais2020rheology}.

Alongside the bulk properties of aggregations, it is important to understand the distribution of individuals inside them. Some of the aggregations we discuss are opaque and several layers thick, so their internal structure cannot be seen with the naked eye. However, they can be investigated with 2D x-ray imaging, as has been done with fire ants \cite{foster2014fire} or with x-ray computed tomography (\ref{g:CT}). Once we know how individuals are arranged, we can  consider them to have distinct liquid crystal phases (\ref{g:LC}). 

This description of the alignment of densely packed rod-shaped particles may prove useful for understanding the behavior of dense invertebrate aggregations of the elongated, yet not entangled organisms. We consider aggregations in which the individuals are not aligned with one another to be examples of a living isotropic (\ref{g:isotropic}) phase. This description applies to piles of fly larvae, wolf spiders, ladybird beetles, and daddy longlegs (\fig{aggregation}{\bf(i,m-p)}) or whirligig beetle clusters  (\fig{aggregation}{\bf(t)}).

Individuals that are aligned with one another through the elongated geometry of their bodies or their motion can be thought of as a nematic phase (\ref{g:nematic}, \ref{g:polar}). A network of \textit{C. elegans worms} in \fig{aggregation}{\bf(s)} is an example of an apolar nematic phase, in which individuals are aligned but travel in opposite directions \cite{sugi2019c}. Examples of a polar nematic phase, in which the individuals are both aligned and travel in the same direction, include confined fly larvae in \fig{bonds}{\bf(c)} and moving piles of sawfly larvae and fungus gnat larvae in \fig{aggregation}{\bf(q,r)}. We can think of aggregations in which individuals are aligned and  arranged in a circle as a cholesteric (\ref{g:cholesteric}) phase. This is often seen in aggregations of sawfly or beetle larvae and is referred to as ``cycloalexy'' \cite{jolivet1990cycloalexy, dury2014preemptive}.

Finally, we can consider invertebrate aggregations in a smectic (\ref{g:smectic}) phase, with individuals arranged in distinct layers. While we do not have many examples of this in the literature, it is a potential description for the honey bees in the chain in \fig{bonds}{\bf(b)} or the processionary caterpillars in \fig{aggregation}{\bf(u,v)}.

Measuring the alignment of individual invertebrates in an aggregation to categorize the aggregations in these phases can help elucidate how the individuals keep their aggregations together through touch receptors on their skin or pheromones \cite{marchetti2013hydrodynamics,bolhuis1997tracing,klamser2018thermodynamic, digregorio2018full,be2020phase}.

Being able to visualize the arrangement of individuals in an aggregation can also allow us to use multi-individual tracking techniques or particle image velocimetry (\ref{g:PIV}) to investigate the motion in these aggregations \cite{pereira2020quantifying, hedrick2008software}. This would improve our understanding of the moving piles of sawfly larvae, fungus gnats, and maggots, and potentially lead to new theories of active matter in three dimensions.

Finally, we can consider the forces experienced by these aggregations. Physical forces between individuals, such as ants pulling on one another with their legs or adhesion via mucus or bristles on the skin of worms, can help determine the structure of the aggregation. We can measure the effects of the weight sensed by an individual, such as army ants in bridges deciding whether to leave the bridge by measuring how many ants walk over them \cite{reid2015army}. Giant honey bees and sawfly larvae are also known to communicate via substrate vibrations, which can be measured with Laser Doppler Vibrometry (\ref{g:LDV}) \cite{kastberger2013social, fletcher2007vibrational}.

Analogies between invertebrate aggregations and living materials could include comparisons to smaller length-scale biological objects, such as cells and tissues. Aggregations in which individuals move in synchronized bursts, such as giant honey bee hives and sawfly larvae, are reminiscent of excitable media, such as spiral waves in the heart \cite{gray1996spiral}. Crawling aggregations, such as piles of fungus gnat larvae, might exert forces on the substrate and one another similarly to cells in a growing tissue \cite{trepat2009physical}, and measuring the substrate forces can provide insight into how these aggregations control the movement of their center of mass.

Going beyond soft matter and fluid mechanics, harnessing methodologies to directly modify the behavior of individual invertebrates can help elucidate their aggregation. For instance, behavior can be manipulated with pharmaceuticals, such as directly feeding octopamine and tyramine to honey bees to decrease their thermoregulatory fanning \cite{cook2017octopamine, kamhi2017origins}. Alternatively, the genes of individuals with known genomes, such as fruit fly larvae, fire ants, and western honey bees, could be modified to increase or suppress aggregation \cite{mckinney2020genetic, wurm2011genome, honeybee2006insights}.

Just as insights from materials science can help study invertebrate aggregations, these aggregations can inspire self-assembled engineered materials of the future. Aggregations of insects and other animals have inspired robot swarms, which also consist of individuals obeying local rules to result in emergent behavior \cite{dorigo2020reflections}. Combining the studies of robot and invertebrate swarms has two potential benefits: first, this provides insight into the behaviors and properties of invertebrates that result in collective behavior, such as modeling entangled blackworms using entangled robots \cite{aydin2020collective}. Second, robotic models of aggregating invertebrates can provide a starting point for future robot swarms or self-healing, active materials. %

In sum, we discussed how aggregations of bees, ants, larvae, worms, and other invertebrates can be understood from a soft matter perspective. The aggregations presented here are only a sample of the blobs, piles, and mass migrations existing in nature. Many other invertebrates form disordered aggregations or piles such as bombardier beetles \cite{schaller2018molecular}, drones of stingless bees on leaves \cite{gruter2020swarming}, dung beetles, carrion beetles, cockroaches, leaf-footed bugs,  juvenile insects, and other arthropods \cite{costa2006other}. We encourage our readers to delve into the countless examples of invertebrate aggregations and hope to inspire a new generation of scientists to explore collective dynamics and bring a deeper understanding of the physics of dense living aggregations.

\clearpage
\section{Glossary}
\label{Glossary}
\begin{enumerate} [label=G\arabic*]

  \item \label{g:active} Active Material - A material in which the constituent particles have an internal source of energy that results in propulsion \cite{marchetti2013hydrodynamics, ramaswamy2010mechanics, ramaswamy2017active}.
  
  \item \label{g:apolar} Apolar particle - An active particle propelled in either direction along its major axis \cite{marchetti2013hydrodynamics, ramaswamy2010mechanics}.
  
  \item \label{g:aspect} Aspect ratio - Ratio of an individual's or an aggregtation's length to its width or height \cite{bolhuis1997tracing}.
  
  \item \label{g:attract} Short range attraction potential - a force that draws individuals closer to one another if they are a short distance away, and does not act if the individuals are farther apart \cite{pham2004glasses}.

  \item \label{g:brownian} Brownian motion - The random motion of a particle in a liquid or gas state \cite{jacobs1935diffusion}.
  
  \item \label{g:cholesteric} Cholesteric - A material state in which the particles are aligned, but with a twist to the structure \cite{tamaoki2001cholesteric}.

  \item \label{g:diffusion} Diffusion - The net movement of a substance from a high chemical gradient to a low chemical gradient \cite{jacobs1935diffusion}.
  
  \item \label{g:dynamical} Dynamical process - the interactions of a group of agents, such as aggregating invertebrates, over time \cite{windeknecht1971general}.
  
  \item \label{g:elasticity} Elasticity - Deformation of a material in response to an applied force that immediately reverses with applied force. An elastic material is also called solidlike \cite{barnes1989introduction}.
  
  \item \label{g:emergent} Emergent behavior - A behavior of a group that arises from the actions and properties of the individuals \cite{corning2002re}.

  \item \label{g:entanglement} Entanglement - Particles in a material connected by interpenetration, such as worms wrapped around one another \cite{gravish2012entangled}. 
  
  \item \label{g:granular} Granular material - a material made out of discrete particles, such as grains of sand or coffee beans \cite{gravish2012entangled}.
  
  \item \label{g:inactive} Inactive material - a material that consists of components without their own energy source.
  
  \item \label{g:isotropic} Isotropic - a material state in which there is no order to the arrangement of the particles, and the values of the material properties are the same in all directions \cite{de1993physics, cladis1975new, bolhuis1997tracing}.
  
  \item \label{g:LC} Liquid crystal - A material that has properties both of liquids and solid crystal, with non-isotropic molecule shapes \cite{de1993physics, cladis1975new, bolhuis1997tracing}.
  
  \item \label{g:LDV} Laser Doppler Vibrometry - a technique that measures the vibrations of a material \cite{kastberger2013social}.

   \item \label{g:MIPS} Motility Induced Phase Separation - Aggregation of active particles into clusters. This is caused by the feedback loop of particles colliding and slowing down, which causes more particles to collide with them and slow down \cite{cates2015motility}. 
  
  \item \label{g:nematic} Nematic - A material state in which the molecules are oriented in the same direction but are not arranged in layers \cite{de1993physics, cladis1975new, bolhuis1997tracing}.
  
  \item \label{g:rheology} Rheology - An experimental technique that measures the viscous and shear properties of a material \cite{barnes1989introduction}.
  
  \item \label{g:order} Order parameter - A measure of the alignment of particles in a material, with 0 being no alignment and 1 meaning all particles are oriented along the same vector \cite{de1993physics}.
  
  \item \label{g:polar} Polar particle - An active particle with a ``head'' and ``tail'' that is propelled in the head-to-tail direction along its major axis \cite{marchetti2013hydrodynamics, ramaswamy2010mechanics}.

  \item \label{g:polarorder} Polar order - particles ordered with their ``heads'' in the same direction \cite{marchetti2013hydrodynamics, ramaswamy2010mechanics}.
  
  \item \label{g:PIV} Particle Image Velocimetry (PIV) - A technique from fluid mechanics that measures the velocity field of a moving fluid by taking consecutive image frames and correlating the positions of particles from frame to frame. The result is a vector field superimposed upon the fluid \cite{thielicke2014pivlab}.
  
  \item \label{g:phase} Phase separation - the spontaneous separation of a material into two or more phases \cite{marchetti2013hydrodynamics}.
  
  \item \label{g:polymer} Polymer - a material made up of long chains of molecules \cite{doi2013soft}.

  \item \label{g:shear} Shear - stress applied tangentially to a material surface \cite{barnes1989introduction}.

  \item \label{g:thin} Shear-thinning - the viscosity of a material decreasing with increasing shear stress \cite{barnes1989introduction}.

  \item \label{g:smectic} Smectic - A material state in which the molecules are oriented in the same direction and are arranged in layers \cite{de1993physics, cladis1975new, bolhuis1997tracing}.
  
  \item \label{g:soft} Soft material - a material that has both solid and liquid properties \cite{doi2013soft}.
  
  \item \label{g:star} Star polymer - a polymer consisting of several polymer chains fused at a central point \cite{ren2016star}.  
  
  \item \label{g:strain} Strain - a stress applied perpendicular to a material surface \cite{barnes1989introduction}.
  
  \item \label{g:surface} Surface tension - the tendency of the surface of a liquid to occupy the smallest possible area caused by the cohesive forces within it.
  
  \item \label{g:viscoelasticity} Viscoelasticity - The combination of viscosity and elasticity that causes some materials to have both viscous (\ref{g:viscosity}) and elastic (\ref{g:elasticity}) components to their response to an applied force. Viscoelastic materials can be thought of as being both solid and liquid \cite{barnes1989introduction}.
  
  \item \label{g:viscosity} Viscosity - Resistance to deformation of a material in response to an applied force caused by friction between its layers. A viscous material can also be thought of as fluidlike \cite{barnes1989introduction}.

  \item \label{g:CT} X-ray Computed Tomography (CT) - Technique that makes a three-dimensional image of all of the features inside an object by taking a set of x-ray projections at angles around an object. Industrial CT refers to scanning large objects in this manner; micro-CT refers to scanning small objects with a high resolution \cite{kalender2006x}.

\end{enumerate}

\section{Supplementary tables}

\begin{table}[h!]
\centering
\begin{tabular}{|c | c | c| c|} 
 \hline
 Behavior & Time scale & Group velocity & Reference \\ [0.5ex] 
 \hline
Giant honey bee shimmering & $<$ 1 s & 0.3 - 0.5 m/s & \cite{kastberger2014speeding} \\ 
 \hline
Sawfly larva twitching & $<$ 1 s & & \cite{boeve2015multimodal, fletcher2007vibrational} \\ 
 \hline
Whirligig beetle flash expansion &  & 0.3 m/s & \cite{boeve2015multimodal, fletcher2007vibrational} \\ 
 \hline
Honey bee swarm flattening & 30 minutes & & \cite{peleg2018collective} \\ 
 \hline
Fire ant raft expansion & 3 minutes & & \cite{mlot2011fire} \\ 
 \hline
Sludge worm phase separation & 60 minutes & & \cite{deblais2020phase} \\ 
 \hline
\textit{C. Elegans} phase separation & 3 minutes & & \cite{sugi2019c} \\ 
 \hline
Fly larva feeding pile & & $10^{-4}$ m/s & \cite{shishkov2019black_1} \\ 
 \hline
Fire ant tower & 20 minutes & $10^{-6}$ m/s & \cite{phonekeo2017fire} \\ 
 \hline
Blackworm blob & & $10^{-5}$ m/s & \cite{aydin2020collective} \\ 
 \hline
Caterpillar procession & & $10^{-3}$ m/s & \cite{uemura2020movement} \\ 
 \hline

\end{tabular}
\caption{Summary of the known time scales and average velocities of aggregation motion.}
\label{veltable}
\end{table}

\section{Acknowledgments}

This work was supported by the National Science Foundation (NSF) Physics of Living Systems Grant No. 2014212 (O.P.). Any opinions, findings, and conclusions or recommendations expressed in this material are those of the authors(s) and do not necessarily reflect the views of the NSF. We also acknowledge funding from the University of Colorado Boulder, BioFrontiers Institute (internal funds), and the Interdisciplinary Research Theme on Multi Functional Materials and Autonomous Systems (O.P.). We thank Jeffery Tomberlin, Chelsea Cook, Raphael Sarfati, Chantal Nguyen, Dieu My Nguyen, Hungtang Ko, Owen Martin, and Golnar Gharooni Fard for reading and commenting on the manuscript.

\bibliographystyle{vancouver}
\bibliography{main.bbl}

\section{Supplementary videos}

\begin{enumerate} [label=S\arabic*]
  \item \label{v:worm} Blackworm (\textit{Lumbriculus variegatus}) blob in water moving across a temperature gradient, 50$^{\circ}$C on the right and 15${^\circ}$C on the left. The aggregated worms move to the preferred colder water, sped up 128x. Video courtesy of Yasemin Ozkan-Aydin and Saad Bhamla.
  \item \label{v:antbridge} Timelapse showing an army ant (\textit{Eciton}) bridge growing as more ants join it. Each frame advances 5 minutes. Video from Ref. \cite{reid2015army}.
  \item \label{v:sawflytwitch} Sawfly larvae (Symphyta) flicking their ``tails'' for defense. Video courtesy of David Yeates.
  \item \label{v:giantbee} Giant honey bees (\textit{Apis dorsata}) on their nest raising their abdomens to form shimmering waves. This behavior wards off approaching wasps. Video from Ref. \cite{kastberger2008social}
  \item \label{v:antraft} A fire ant (\textit{Solenopsis invicta} aggregation consisting of 8,000 ants spreading out from a ball placed in water to a flat raft. Video sped up 15x, from Ref. \cite{mlot2011fire}.
  \item \label{v:blowfly} A migration of blow fly larvae from a carcass they consumed. The pile of larvae spreads out and travels in the same direction. Video from Ref. \cite{lashley2018indirect}. 
  \item \label{v:fungusgnat} A ``migration'' of fungus gnat larvae (Sciaridae) making their way across a yard in Finland. Video by Olli Korhonen.
  \item \label{v:sawflymig} A ``migration'' of sawfly larvae (Symphyta) near the summit of Mount Coot-Tha in Brisbane. Video by Benjamin R. Fitzpatrick.
  \item \label{v:onecaterpillar} A procession of caterpillars (\textit{Thaumetopoea pityocampa}) searching for a pupation site. Video from Ref. \cite{uemura2020movement}.
  \item \label{v:twocaterpillars} Two caterpillar species, \textit{Hylesia metabus} and \textit{Hylesia nanus}, aggregating into a procession in a circular dish and then walking around, sped up 64x. Video courtesy of Stefanie White and Amy Deacon.
\end{enumerate}

\end{document}